\documentclass[conference]{IEEEtran}
\IEEEoverridecommandlockouts

\usepackage{cite}
\usepackage{amsmath,amssymb,amsfonts}
\usepackage{algorithmic}
\usepackage{graphicx}
\usepackage{textcomp}
\usepackage{xcolor}
\def\BibTeX{{\rm B\kern-.05em{\sc i\kern-.025em b}\kern-.08em
    T\kern-.1667em\lower.7ex\hbox{E}\kern-.125emX}}

\usepackage{fancyhdr}
\usepackage{hyperref} 
\hypersetup{hidelinks}

\fancypagestyle{ieee}{ 
    \fancyhf{} 
    \fancyhead[C]{
    \footnotesize This work has been accepted for publication in 2025 37th International Conference on Microelectronics (ICM). This is the author’s version which has not been fully edited and content may change prior to final publication. Citation information: DOI 10.1109/ICM66518.2025.11322444}
    \fancyfoot[C]{
    \footnotesize \copyright~2025 IEEE.  Personal use of this material is permitted.  Permission from IEEE must be obtained for all other uses, in any current or future media, including reprinting/republishing this material for advertising or promotional purposes, creating new collective works, for resale or redistribution to servers or lists, or reuse of any copyrighted component of this work in other works.}

}

\begin{document}

\title{DISCA: A Digital In-memory Stochastic Computing Architecture Using A Compressed Bent-Pyramid Format}

\author{
\IEEEauthorblockN{Shady Agwa\IEEEauthorrefmark{1},
Yikang Shen\IEEEauthorrefmark{2},
Shiwei Wang\IEEEauthorrefmark{1},
Themis Prodromakis\IEEEauthorrefmark{1}}
\IEEEauthorblockA{\IEEEauthorrefmark{1}\textit{Centre for Electronics Frontiers, Institute for Integrated Micro and Nano Systems,}\\
\textit{School of Engineering, The University of Edinburgh, UK.}\\
shady.agwa@ed.ac.uk, shiwei.wang@ed.ac.uk, t.prodromakis@ed.ac.uk}
\IEEEauthorblockA{\IEEEauthorrefmark{2}\textit{Electrical \& Electronics Engineering,} 
\textit{Queen's University Belfast, UK.}\\
yshen19@qub.ac.uk}
\thanks{\textit{\IEEEauthorrefmark{2}Yikang Shen contributed to this work while he was at the School of Engineering, The University of Edinburgh, UK.} }
}
\maketitle
\thispagestyle{ieee} 

\begin{abstract}
Nowadays, we are witnessing an Artificial Intelligence revolution that dominates the technology landscape in various application domains, such as healthcare, robotics, automotive, security, and defense. Massive-scale AI models, which mimic the human brain’s functionality, typically feature millions and even billions of parameters through data-intensive matrix multiplication tasks. While conventional Von-Neumann architectures struggle with the memory wall and the end of Moore’s Law, these AI applications are migrating rapidly towards the edge, such as in robotics and unmanned aerial vehicles for surveillance, thereby adding more constraints to the hardware budget of AI architectures at the edge. Although in-memory computing has been proposed as a promising solution for the memory wall, both analog and digital in-memory computing architectures suffer from substantial degradation of the proposed benefits due to various design limitations. We propose a new digital in-memory stochastic computing architecture, DISCA, utilizing a compressed version of the quasi-stochastic Bent-Pyramid data format. DISCA inherits the same computational simplicity of analog computing, while preserving the same scalability, productivity, and reliability of digital systems. Post-layout modeling results of DISCA show an energy efficiency of 3.59 TOPS/W per bit at 500 MHz using a commercial 180nm CMOS technology. Therefore, DISCA significantly improves the energy efficiency for matrix multiplication workloads by orders of magnitude if scaled and compared to its counterpart architectures.

\end{abstract}
\section{Introduction}
The main computational core of the emerging AI models is the Matrix-Matrix Multiplication ($MatMul$). 
Therefore, AI architectures have to perform millions, and even billions, of Multiply-Accumulate (MAC) operations. 
This adds more performance and energy-efficiency demands to the classical Von Neumann computing architectures, which face major challenges due to the 
data-intensive/compute-intensive nature of these emerging massive-scale AI models. 
A major challenge is the high cost of data movement between memory and processing units 
as a result of the Von Neumann bottleneck. Another challenge is the computational complexity of 
$MatMul$ in the binary computing domain, requiring massive hardware resources, which are not energy efficient.

Analog In-Memory Computing (IMC) architectures have been widely proposed to overcome the Von Neumann bottleneck using emerging technologies, such as RRAMs~\cite{Yao_nature2020,Wan_nature2022,Kim_jetcas2022}. 
Analog IMC depends on resistive crossbar arrays to perform $MatMul$ workloads in the analog domain 
through Ohm's law and Kirchhoff's current law. Although the multiplication and accumulation operations are 
easily performed, analog IMC architectures still inherit the same 
complexity challenges of the analog design~\cite{Adam_nature2018,Liu_isscc2020}. The first 
challenge is the high variability of analog memristors, including device-to-device variations and cycle-to-cycle variations. Furthermore, these emerging devices face resistance drifting over time, which requires continuous calibration. 
This puts more constraints on the writing and reading circuitry. The second major challenge is 
the analog/digital cross-domain interfacing complexity, where Digital-to-Analog Converters (DACs) and Analog-to-Digital Converters (ADCs) are required with good accuracy. Although analog IMC architectures can run at very 
low supply voltage to achieve high energy efficiency, the current-based computation and the 
complexity of the analog/digital cross-domain interfacing lead to long latencies and low throughput, 
which are not suitable for AI applications. Analog IMC also suffers from low computational/memory density 
due to the high complexity of the peripheral circuitry. Eventually, analog IMC approaches are still struggling with design scalability from system-size and technology node perspectives; meanwhile, AI models have been significantly scaled up to trillions of parameters since the Transformer era started.

In another direction, digital IMC architectures are also proposed to minimize data movement 
while inheriting the same productivity and scalability of the digital design approach. 
These digital architectures leverage the massive parallel resources of the memory structure to 
process data where it exists in DRAMs~\cite{seshadri-micro2013,Farahani-hpca2015}, RRAMs~\cite{agwa_iscas2022}, or 
SRAMs~\cite{Jeloka-vlsic2015,Eckert-ISCA2018,Fujiki-ISCA2019,Al-Hawaj-iscas2020,Al-Hawaj-hpca2023} using a bitline computing approach. The bitline computing approach depends on activating two memory wordlines simultaneously
to achieve bitwise logical operations (AND and NOR) between the two wordlines during the read operation. 
Then, a near-memory logic stack is tightly coupled with the memory peripherals to perform more 
complex operations, such as multiplications and additions, 
but at the cost of both throughput and energy efficiency. 
For example, digital IMC consumes up to hundreds of cycles to perform MAC operations due to their complex micro-algorithms~\cite{Al-Hawaj-iscas2020}. Thus, digital binary-based IMC architectures struggle to meet the high-throughput and energy-efficiency expectations.

Stochastic Computing (SC)~\cite{Alaghi_acm2013,Alaghi_dac2013,Alaghi_date2014,Winstead_springer2019,Groszewski_isqed2019,Alaghi_phd2015,Zhang_snw2019,Salehi_tvlsi2020,Alaghi_tcadics2018,Zhang_tcasii2020} 
was proposed as a middle ground between digital (binary-based) and analog computing domains. 
It inherits the computational simplicity of the analog domain while still keeping the same 
scalability, reliability, and productivity of the digital domain.

In this paper, we propose a Digital In-SRAM Stochastic Computing Architecture (DISCA) for $MatMul$ workloads. This novel architecture leverages the bitline computing approach to perform Stochastic Multiplication (SC-MUL) operations and accumulation periphery, including parallel counters and adder trees to achieve the $MatMul$ functionality. DISCA leverages conventional CMOS technology nodes, thereby avoiding any uncertainties or scaling challenges associated with emerging devices, such as RRAMs. As a showcase, DISCA utilizes a compressed version (8-bit) of the Bent-Pyramid (BP) data format, which is a quasi-stochastic data representation that avoids the primary pitfalls of conventional stochastic bitstreams~\cite{agwa_newcas23,oisma_2025}. Using a commercial 180nm CMOS technology node, the hardware implementation results show that DISCA operates at a 500 MHz clock Frequency, achieving an energy efficiency of 3.59 TOPS/W per bit. These results show better energy efficiency over digital IMC architectures, implemented by a five-generation 
advanced technology node~\cite{Al-Hawaj-iscas2020}. Scaling down the technology of DISCA to equivalent advanced nodes results in an average of two orders of magnitude improvement in the energy efficiency~\cite{Stillmaker_integration2017}.

 The paper is organized as follows: Section II discusses the basics of stochastic computing and the compressed version of Bent-Pyramid. In Section III, DISCA architecture is introduced. Section IV presents the hardware implementation results. Finally, Section V is the conclusion and the future directions.
\section{Compressed Bent-Pyramid}
\begin{figure}
\includegraphics[width=0.9\linewidth]{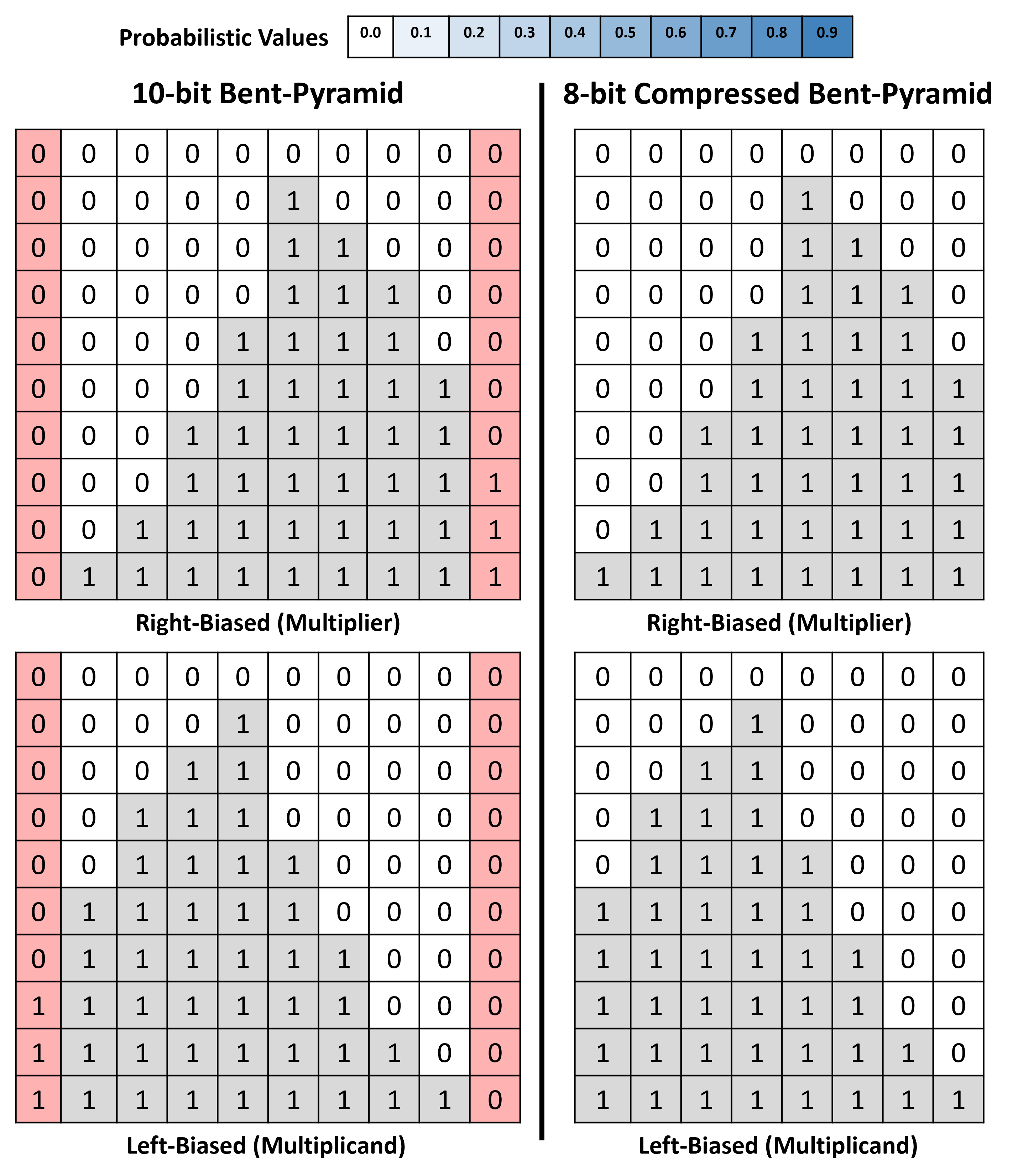}
\centering
\caption{A compressed 8-bit Version of Bent-Pyramid datasets, achieving the same 10-bit Bent-Pyramid functionality.}
\label{cbp}
\end{figure}
Conventional SC numbers are represented by randomly-generated bitstreams of ones and zeros, where bitstreams are mapped to probabilities (from 0.0 to 1.0). Regarding the unipolar stochastic number, which is the focus of this work, the equivalent probability is calculated by the ratio of ones to the total number of bits in the SC bitstream~\cite{Alaghi_phd2015}. For unipolar SC bitstreams, the SC multiplication is achieved by a bitwise logical AND operation. This multiplication simplicity has been recognized as a promising computing domain for AI applications~\cite{Zhang_tcasii2020,agwa_frontiers2023,agwa_newcas23}.

Typically, an n-bit binary number $B$ is converted to a $2^n$-bit SC number $S$, to achieve full precision mapping accuracy. 
This n-bit binary B is compared to an n-bit pseudo-random number $R$, which is generated by an n-bit Linear Feedback Shift Register (LFSR). After $2^n$ cycles, the output bitstream $S$ represents the probability P\textsubscript{S} of $B/2^n$. These longer bitstreams and latencies of SC number generation are the main pitfalls of the SC computing domain.
A recent work~\cite{agwa_frontiers2023} targeted improving the accuracy of SC computing for a Vector-Matrix Multiplication (VMM) benchmarking throughout three different phases: 1) SC number generation, 2) SC multiplication, and 3) SC numbers accumulation. The accuracy analysis shows that using quasi-stochastic numbers can map the n-bit binary domain to lower than $2^n$ stochastic bit-precisions with minimal error. By utilizing a fully binary accumulation approach, the accuracy results of the 4-bit binary VMM benchmark show a negligible average error of less than 1\% at 4-bit SC precision. 
To avoid the SC generation bottleneck and the indeterministic computational behavior of SC bitstreams, Bent-Pyramid (BP)~\cite{agwa_newcas23} was proposed as a quasi-stochastic data format that can replace the conventional SC bitstreams, with an average error of 2.13\% for VMM benchmarking. The BP system has two complementary bitstream datasets of 10 bits, representing probabilities from 0.0 to 0.9 with
10\% step resolution. 
A compressed version of the 10-bit BP format was proposed to achieve the same functionality while utilizing only 8 bits (BP8)~\cite{oisma_2025}. Figure~\ref{cbp} illustrates the typical 10-bit and compressed 8-bit BP systems. The most left and right bits are eliminated for both left and right-biased complementary datasets, leading to BP8 with two extra implicit bits that are not physically implemented by hardware. Various $MatMul$ benchmarking workloads show a marginal accuracy loss for BP8 in comparison to floating points~\cite{oisma_2025}. BP8 provides approximate $MatMul$ computations; however, these computations are deterministic, therefore the same inputs always lead to the same outputs, while any approximation can be tolerated by re-adjusting the pretrained weights of the AI models.

In this paper, DISCA architecture features the quasi-stochastic multiplication computing and fully binary accumulation of BP8 format, targeting $MatMul$ workloads for AI.

\section{DISCA Architecture}
\begin{figure*}
\includegraphics[width=0.9\linewidth]{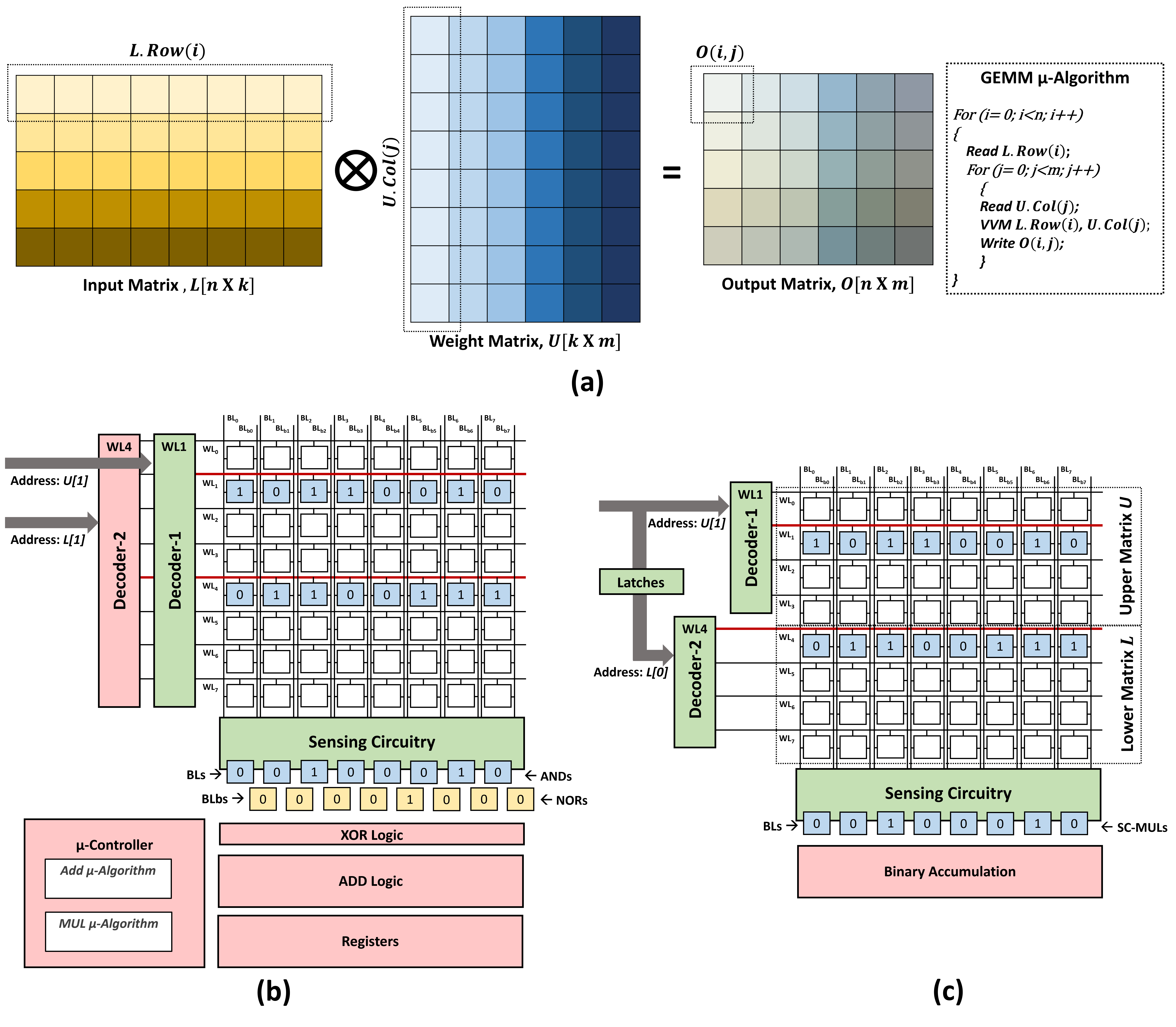}
\centering
\caption{In-memory computing for matrix-matrix multiplication: (a) an example of matrix-matrix multiplication workload including the micro-algorithm, where input matrix $L$ is multiplied by weight matrix $U$ to generate the output matrix $O$; (b) digital in-memory computing architecture for binary-based computing; (c) the proposed DISCA architecture for matrix-matrix multiplication using in-situ stochastic multiplication and near-memory binary accumulation.}
\label{GEMM_BIMC_vs_SIMC}
\end{figure*}

Figure~\ref{GEMM_BIMC_vs_SIMC}(a) shows an illustrating example of a $MatMul$ workload, where the input matrix $L$ is multiplied by the weight matrix $U$ to generate the output matrix $O$. A micro-algorithm is also shown for the computing of the $MatMul$, where a row of values from matrix $L$ is read and then multiple rows of matrix $U$ are read sequentially. Each new row of matrix $U$ will be multiplied by the row from matrix $L$ to calculate an output value of matrix $O$. Then, another row from matrix $L$ will be read, and the above process should be repeated till achieving all results of matrix $O$.

To achieve the $MatMul$ operations through a conventional digital IMC, data from the input matrix $L$ and weight matrix $U$ should be aligned within the same memory sub-array, as shown by Figure~\ref{GEMM_BIMC_vs_SIMC}(b). Two decoders are required to activate two wordlines simultaneously: one wordline for the upper weight matrix $U$ and another wordline for the lower input matrix $L$. The in-situ processing provides bitwise AND and NOR operations. A compute-logic stack utilizes these two operations to achieve more complex operations, such as additions. A microcontroller runs different micro-algorithms to perform multiplication and accumulation (MAC) operations, which can consume hundreds of cycles. Consequently, digital binary-based IMC suffers from long latencies and considerable degradation in performance and energy efficiency.

In contrast, DISCA architecture adopts a quasi-stochastic computing domain with a binary accumulation approach. Figure~\ref{GEMM_BIMC_vs_SIMC}(c) shows that DISCA utilizes only bitwise AND operations to achieve unipolar SC multiplications. To avoid using two decoders, the data layout of the two matrices follows specific constraints, where the lower memory part hosts the input matrix $L$ and the upper memory part is dedicated to the weight matrix $U$. Assuming that the memory is divided equally between $L$ and $U$, the hierarchical decoder ($n~to~2^n$) is split into two smaller decoders ($n~to~2^{n-1}$) with a latching option for the lower decoder's address. Thus, the first address is received by the two small decoders to activate two wordlines simultaneously, e.g., $L.Row(0)$ equivalent to $MEM.Row(4)$ and $U.Row(0)$ equivalent to $MEM.Row(0)$. Then, a bitline computing operation takes place to achieve bitwise AND operations. For the next round of SC multiplications, the latches are disabled to keep the same $L.Row(0)$ activated and a new address propagates to the upper memory decoder to activate a new wordline of the weight matrix, e.g., $U.Row(1)$ equivalent to $MEM.Row(1)$ as shown by Figure~\ref{GEMM_BIMC_vs_SIMC}(c).
This shared decoding logic allows all wordlines of matrix $U$ to be multiplied by one wordline of matrix $L$ before moving to the next wordline of matrix $L$. Each row of matrix $L$, shown in Figure~\ref{GEMM_BIMC_vs_SIMC}(a), is mapped as a memory wordline in Figure~\ref{GEMM_BIMC_vs_SIMC}(c), while each column of matrix $U$, shown in Figure~\ref{GEMM_BIMC_vs_SIMC}(a), is mapped as a memory wordline in Figure~\ref{GEMM_BIMC_vs_SIMC}(c).

DISCA proposes reducing the hardware overhead of digital IMC by removing the extra decoder and microcontroller, while replacing the complex compute logic stack with a binary accumulation periphery. This binary accumulation logic handles the SC to binary conversion on-the-fly while performing the accumulation of $MatMul$ workloads.

DISCA inherits the same scalability of on-chip memories, aiming to achieve the same scale of Last Level Caches (LLCs); therefore, providing massively parallel computing resources for $MatMul$ workloads. For example, a 128KB DISCA engine can feature eight DISCA banks, and each bank consists of four 4KB subarrays. The 4KB DISCA subarray typically has 256 columns (C) and 128 rows (R) of 6T CMOS SRAM bitcells.

\section{Hardware Implementation}
\begin{figure}
\includegraphics[width=1\linewidth]{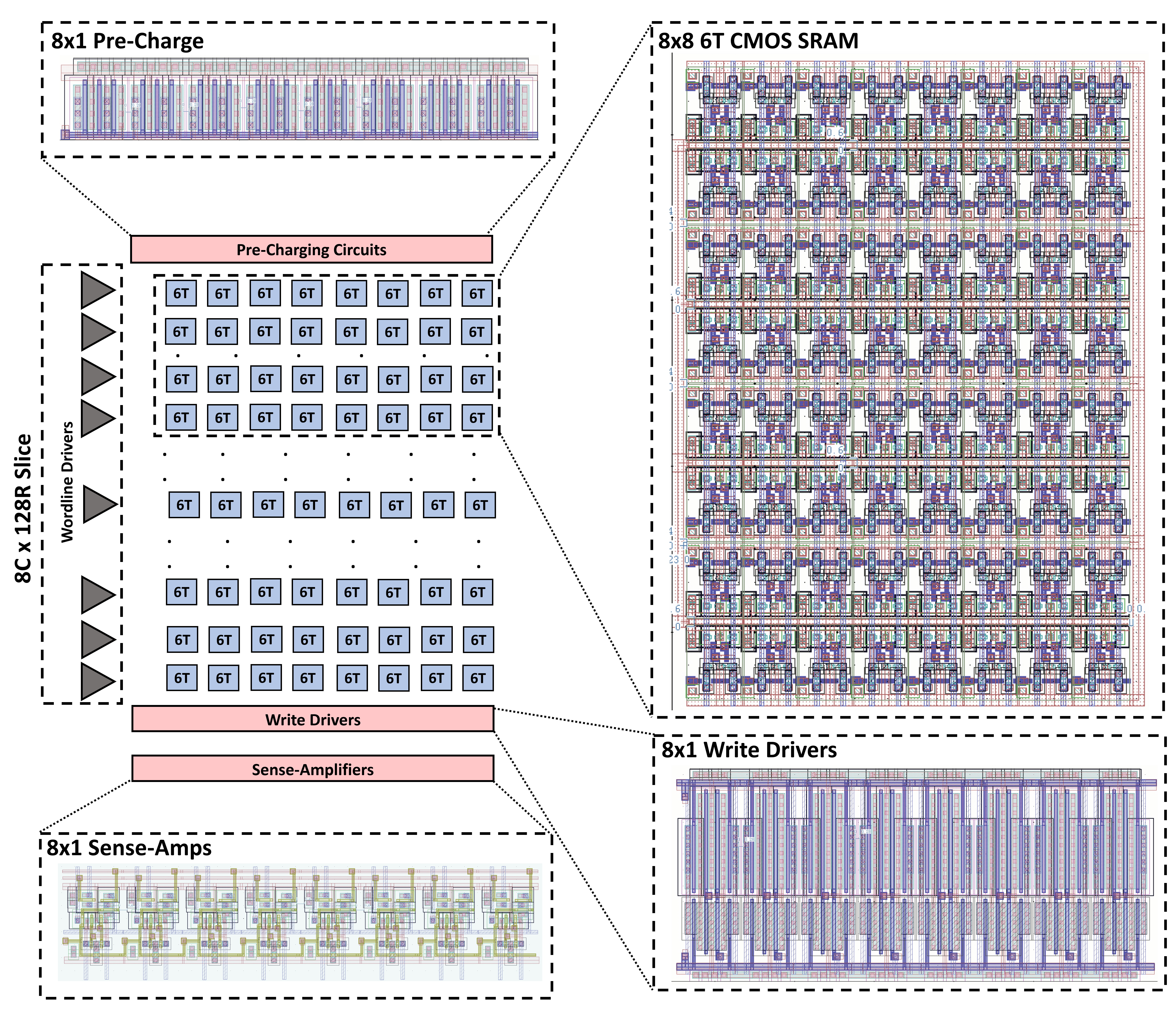}
\centering
\caption{An 8CX128R SRAM core slice, including the layout of 8x8 6T bitcells macro, pre-charge circuitry, write drivers, and sense-amplifiers.}
\label{Slice_Layout}
\end{figure}
\begin{figure}
\includegraphics[width=1\linewidth]{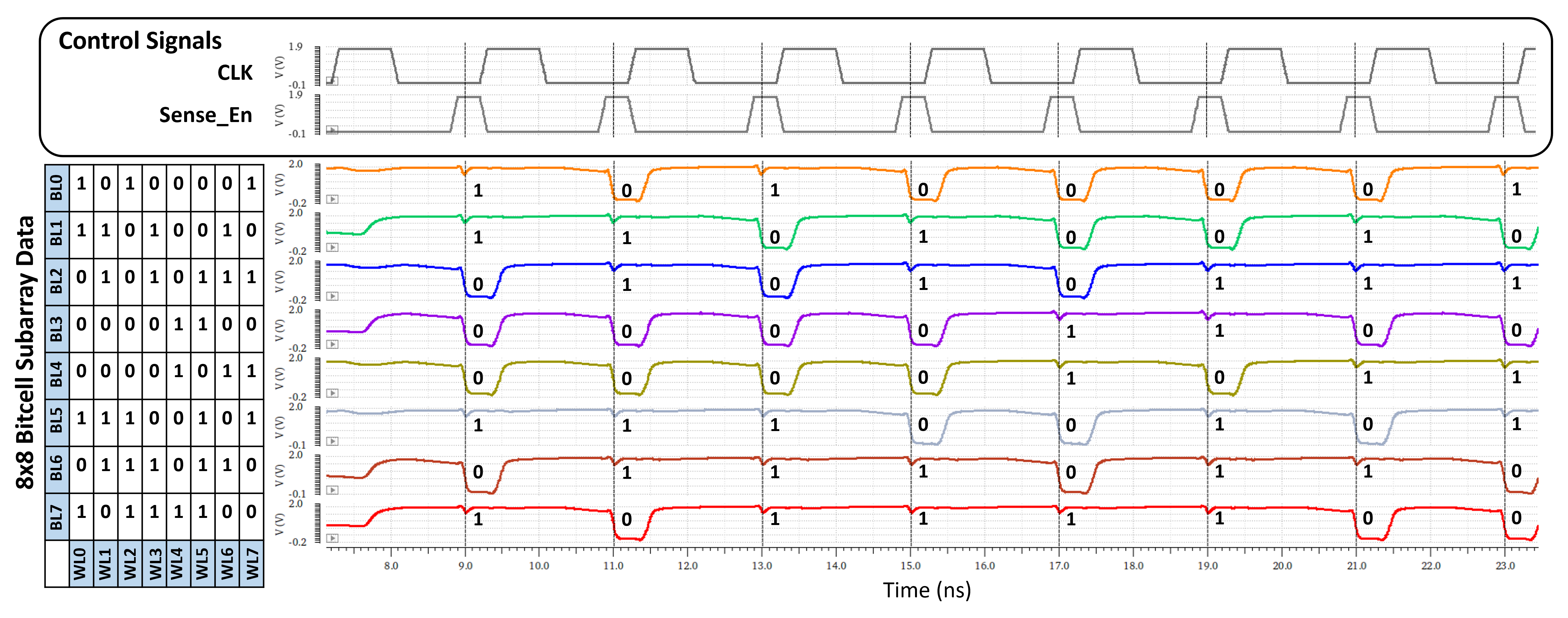}
\centering
\caption{Simulations of eight read operations for different 8-bit values stored in the SRAM core.}
\label{sim_read}
\end{figure}
\begin{figure}
\includegraphics[width=1\linewidth]{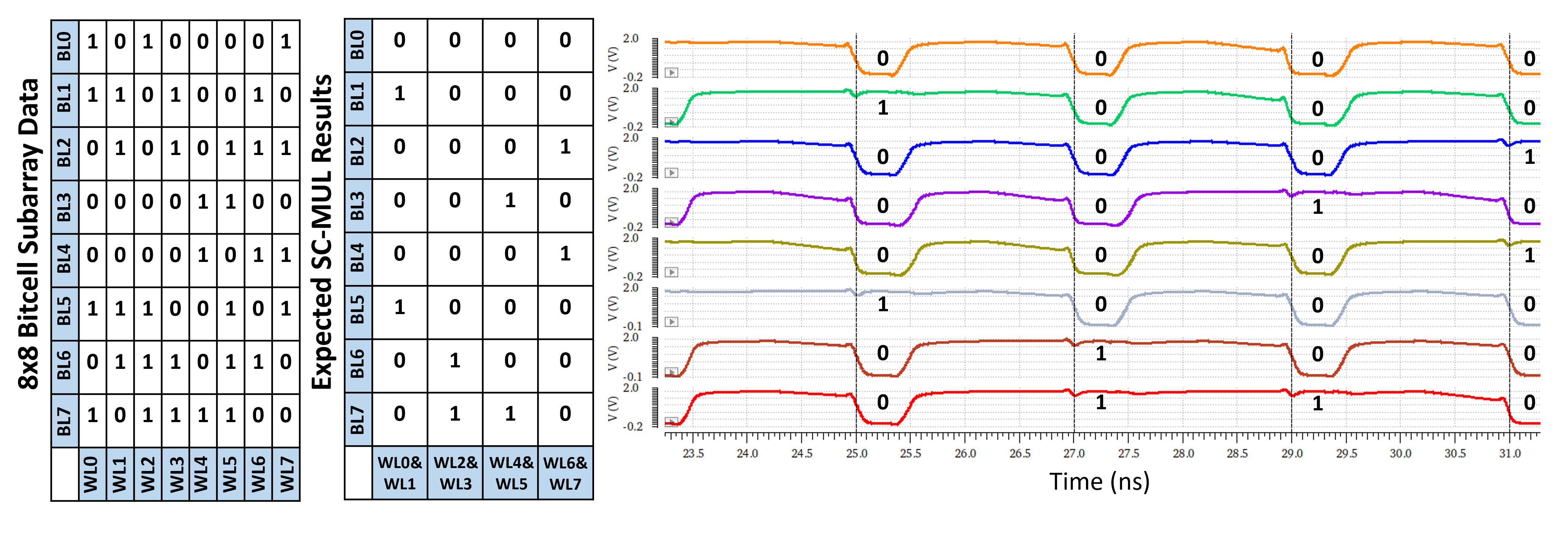}
\centering
\caption{Simulation results of SC multiplication using the bitline computing approach where the upper matrix $U$ and lower matrix $L$ are logically represented by even wordlines and odd wordlines respectively.}
\label{sim_scmul}
\end{figure}

To model the proposed DISCA architecture, a digital Cell-based ASIC design flow was used to implement the address decoder and the SC-to-Binary accumulation periphery; The RTL was synthesized using Cadence, Genus and the Place-\&-Route was done using Cadence Innovus. Meanwhile, a custom IC design flow was used to implement the 6T SRAM memory core. The transistor-level implementation was accomplished to the layout using Cadence Virtuoso. 
The energy and performance results were extracted Post-layout and Place-\&-Route for the different components, using commercial 180nm CMOS technology node.
\subsection{SC-to-Binary Accumulator}
The SC-to-Binary Accumulator incorporates a layer of parallel counters followed by an adder tree to accumulate a full wordline of SC multiplication results (256 bits), following the same approach in ~\cite{oisma_2025}. The 256-bit stochastic bitstream is programmable to incorporate $n$ SC multiplications of $m$-bit bitstreams, where $nxm$ equals 256 bits. Assuming that DISCA adapts the compressed 8-bit BP format, the SC-to-Binary Accumulator accumulates 32 different  SC multiplications per cycle. 
Both parallel counters and adder trees were implemented using Verilog and a commercial 180nm CMOS technology node. A pipeline stage was inserted to split the SC-to-Binary Accumulator into two stages for better timing. The post-Place \& Route results show that the SC-to-Binary Accumulator consumes an average power of 31.14 mW at 500 MHz, leading to 243.28 fJ/bit for SC-to-binary conversion and accumulation.

\subsection{DISCA Decoder}
The DISCA decoder was also implemented using Verilog and the same commercial 180nm CMOS technology node. The post-Place \& Route results show an average power consumption of 8.415 mW at 1 GHz, where the decoding period is half the period of the memory read or bitline compute operations. Thanks to the data regularity and massive parallelism of $MatMul$ workloads, the same decoder can be shared by two DISCA subarrays simultaneously, therefore reducing the decoding energy cost to 13.15 fJ/bit.

\subsection{SRAM Core}

To build a 256Cx128R SRAM core with bitline computing capability, a macro of 8x8 6T CMOS SRAM bitcells was implemented in the full layout, using a commercial 180nm CMOS technology node. Other peripheral circuits, such as pre-charge circuits, write drivers, and sense amplifiers, were also implemented. To enable the bitline computing functionality, the sense-amplifier was modified to be reconfigurable. The proposed sense-amplifier has two modes of operation: 1) Differential mode, where it receives both bitline $BL$ and bitline-bar $BLb$, and 2) single-ended mode, where it receives only bitline $BL$ and compares it against a reference voltage $V_{ref}$.

Figure~\ref{Slice_Layout} shows an 8CX128R SRAM slice, including the layout of 8x8 6T bitcell macro, pre-charge circuitry, write drivers, and sense-amplifiers. All blocks were extracted and then post-layout simulated. 
The functionality of the SRAM slice was verified for write, read, and SC multiplication operations using Cadence Virtuoso. Figure~\ref{sim_read} shows the simulation results of eight read operations for 8-bit values. These values were previously written successfully to the SRAM slice, considering balanced values of 0s and 1s for representative average energy results. Figure~\ref{sim_scmul} illustrates the simulation results of SC multiplication using the bitline computing approach. 
The energy consumption results at 500 MHz are 411 fJ/bit for the write operation, 289.75 fJ/bit for the read operation, and 301 fJ/bit for the SC multiplication operation. The results show that SC multiplication consumes slightly higher energy than its read counterpart (3.9\%), due to activating two wordlines simultaneously. It is also worth mentioning that DISCA read operation consumes higher energy by 4.5\% in comparison to its equivalent SRAM without bitline computing capability.

\subsection{DISCA Engine}
From the previous hardware energy and performance results, an accurate model can be exploited for the proposed DISCA Engine (e.g., 128KB). The total energy consumption of SC multiplication, including shared decoders, is 314.15 fJ/bit at 500 MHz.

Considering the results of SC-to-Binary Accumulator, DISCA MAC operation consumes 557.428 fJ/bit for the hybrid (SC/binary) multiplication/accumulation. Consequently, the energy efficiency of DISCA is 3.59 TOPS/W per bit, while the peak throughput of the 128KB DISCA Engine is 7.9 TOPS per bit, counting two operations per MAC. Regarding the compressed BP8 format, DISCA has an energy efficiency of 448.5 GOPS/W and throughput of 0.988 TOPS.

Scaling DISCA to 22nm technology using DeepScaleTool\cite{stillmaker2017scaling}\cite{sarangi2021deepscale}, increases the energy efficiency by two orders of magnitude, enabling DISCA to outperform its digital binary-based IMC counterparts by orders of magnitude, such as \cite{Al-Hawaj-iscas2020} and \cite{eidetic2023}.

\section{Conclusion}

In this work, we propose a new Digital In-SRAM Stochastic Computing Architecture (DISCA) for $MatMul$ workloads, utilizing a bitline computing approach to perform massively parallel stochastic multiplication operations. A near-memory logic, including parallel counters and adder trees, receives the output multiplication bitstreams for binary accumulation to achieve the $MatMul$ functionality. DISCA achieves an energy efficiency of 3.59 TOPS/W per bit using a commercial 180nm CMOS technology node. This highlights a strong promise in addressing the increasing performance and energy-efficiency demands of modern AI workloads. DISCA is a flexible architecture that can adapt to various stochastic bitstream lengths, leading to a promising trade-off between accuracy and energy efficiency. In this work, DISCA features a compressed version of the Bent-Pyramid format (BP8) as a showcase, therefore avoiding the primary pitfalls of conventional stochastic bitstreams. DISCA provides an energy efficiency of 448.5 GOPS/W using the 8-bit BP format, outperforming digital binary-based IMC architectures by orders of magnitude, if scaled to the same technology node. Moreover, DISCA leverages conventional CMOS technology nodes, thereby avoiding the uncertainties and scaling challenges of the emerging devices, such as RRAMs.

The future work will focus on scaling up DISCA to build a multi-array engine while scaling down the implementation technology to 22nm. In another direction, further investigations are required to improve the code density of the BP system, thereby increasing the overall computational and memory density.
\section*{Acknowledgment}

\thanks{This work was supported by EPSRC
FORTE Programme (Grant No. EP/R024642/2) and by the RAEng Chair in Emerging Technologies (Grant No. CiET1819/2/93).}

\end{document}